\documentclass[final,5p,times,twocolumn]{elsarticle}

\usepackage{amssymb}
\usepackage{xcolor}
 \usepackage{amsmath}
\usepackage{rotating}
\usepackage{siunitx}

\newcommand*{\citen}{}
\DeclareRobustCommand*{\citen}[1]{%
  \begingroup
    \romannumeral-`\x 
    \setcitestyle{numbers}%
    \cite{#1}%
  \endgroup
}

\journal{Computational Materials Science}

\begin{document}

\begin{frontmatter}



\title{Ground state determination and band gaps of bilayers of graphenylenes and octafunctionalized-biphenylenes}

\author[label1]{Chad E. Junkermeier}
\ead{junkerme@hawaii.edu}
\author[label2]{Ricardo Paupitz}
\address[label1]{STEM Department, University of Hawai`i Maui College, Kahului HI 96732, USA}
\address[label2]{Departamento de F\'{\i}sica, IGCE, Universidade Estadual Paulista, UNESP, 13506-900, Rio Claro, SP, Brazil}

\begin{abstract}
Device fabrication often requires materials that are either reliably conducting, reliably semiconducting, or reliably non-conducting. Bilayer graphene (BLG) changes from a superconductor \cite{cao2018unconventional} to a semiconductor \cite{ohta2006controlling} depending on it's stacking, but because it is difficult to control its stacking, it is not a reliable material for device fabrication \cite{Bistritzer2011Moiré, Kim3364} . Using DFTB+\cite{aradi2007dftb}, this work demonstrates that bilayers of graphenylene, net-C, and net-W can be reliably used for device fabrication without knowing the details of their stackings. Bilayers of graphenylene and net-C are semiconducting for all sheer displacements, net-W is conducting for all sheer displacements, while that Type II, like BLG, is conducting or semiconducting depending on the sheer displacement.  The method used gives bond lengths, unit cell dimensions, and band dispersion of single-layer graphene that are consistent with previously reported values, it correctly predicts that AB stacking is the ground state of BLG and gives an interlayer separation that is consistent with previous studies. The bond lengths and lattice constants of the other carbon allotropes are consistent with previously published values. In order to calculate the band structures the bilayer systems, DFTB+ was first used to determined the interlayer separations of the 2-D carbon allotropes under shear displacement.
\end{abstract}


\end{frontmatter}


\section{Introduction}
The ground state of carbon is graphite; a layered solid made up of single-atom-thick sheets of carbon held together by dispersive forces.  The sheets, called graphene, are formed by carbon atoms each in an sp2 valence electron configuration, that interlink into a hexagonal pattern, shown in Figure \ref{fig:organiccells} (a), via sigma and pi-pi bonds (often denoted together as pi-bonds). The first modern, reproducible synthesis of graphene as a new and separate allotrope of carbon was experimentally discovered in the early part of this century\cite{Novoselov04666}.
Because the gapless nature of graphene is troublesome for specific applications, methods to create a band gap have been investigated\cite{Robinson121990,fluoro2013,bieri2009porous,gui2008band,panchakarla2009synthesis,han2007energy,areshkin2007building}.  One method for opening a band gap is by applying an electric field across bilayer graphene (BLG) \cite{Novoselov04666}.
Stacking sequence and twist angle play essential roles in the properties of BLG \cite{wang2012interfacial}.
For example, the electronic structures of the layers decouple in twisted BLG for twist angles $>$ \ang{10}, while BLG can become superconducting at so-called magic angles \cite{SuarezMorell2010121407,Bistritzer2011Moiré, cao2018unconventional}.
Thus, while graphene, and by extension BLG, is a promising material other similar materials might be better suited to industrial processing.

Graphenylene, also called biphenylene-carbon (BPC) \cite{Brunetto1212810}, was recently synthesized using the chemical precursor 1,3,5-trihydroxybenzene \cite{du1740796}. It is in the P6/mmm space group \cite{Song1338}, similar to graphene, but with the C6 rings in the place of graphene's carbon atoms with bonds between unit cells formed by the creation of C4 rings, as can be seen in Figure \ref{fig:organiccells} (b). Linking of the biphenylene groups forms C12 rings.  Molecular dynamics (MD) simulations demonstrate that BPC based structures are stable up to thousands of degrees\cite{Rahaman201765, C4CP03529A}.
Three other biphenylene based carbon allotropes, each with 4-, 6-, and 8-member rings (C8) have also been proposed.
The first, net-C \cite{tyutyulkov1997structure}, represented in Figure \ref{fig:organiccells} (c) is in the space group Pmmm.
Simulated IR spectrum of net-C finds that all vibrational wave numbers are real, indicating the allotrope is stable~\cite{Karaush2014229}. Computed band structures of net-C range from metallic to a band gap of 2 eV \cite{tyutyulkov1997structure, hudspeth2010electronic, denis2014stability}.
The 2-D allotrope in \ref{fig:organiccells} (d), space group Pm, has only been discussed in one paper of which we know, and then just long enough to be given the moniker ``Type II" \cite{ANGE:ANGE201309324}.
Net-W, represented in Figure \ref{fig:organiccells} (e), is in space group Cmmm \cite{wang2013prediction}.
Phonon-mode analysis and MD simulations at 300 K suggest that net-W is structurally stable \cite{wang2013prediction}.
One previously reported net-W band structure report results that suggest it is metallic in nature \cite{wang2013prediction}, while another indicates that it is a zero-gap semiconductor \cite{Xu141113}.  These structures, when studied, are often considered for use cases involving lithium storage \cite{yu2013graphenylene,Denis201530} and gas separation \cite{Song1338}.

This work aimed to determine the ground state stacking configurations of graphenylene, net-C, Type II, and net-W.  It gives the relative energies, interlayer spacings, and band gaps change as one sheet is displaced with respect to another. The results in this work compare well to previously published values.  It also discusses the minimum energy displacement paths between pairs of ground state configurations.  These results demonstrate that compared with graphene the energy cost of sheering motions is much higher.  An accurate description of the ground state configurations of theses bilayer structures has not yet been discussed in the literature.  This work provides an essential first step into discussing a range of new possibilities for devices constructed from these materials.

\section{Methodology}

The structures studied in this work are the bilayer analogs of graphene, graphenylene, net-C, Type II, and net-W.  Geometry optimizations of the single layer structures were performed using the density functional based tight binding (DFTB) method, implemented in DFTB+ \cite{Elstner1998,aradi2007dftb,manzano2012}. DFTB+ has near density functional theory (DFT) precision in electronic structure calculations while being significantly faster than DFT. It is based on a second-order expansion of the Kohn-Sham energy defined in DFT \cite{Elstner1998,Kubar2013}. In the present work, we used a particular parametrization, intended to describe systems of interest for materials science,
as implemented in the {\sl matsci Slater-Koster files} \cite{frenzel2004semi,lukose2010reticular}. Dispersion terms were included through the use of Lennard-Jones potentials with the requisite parameters taken from the Universal Force Field (UFF) \cite{Rappe1992UFF,LEBEDEVA201745, Shin2017interlayer}.  For geometry optimization, a conjugate gradient algorithm was used imposing a maximum force difference of $10^{-5}$ and a maximum tolerance of $10^{-4}$ as convergence criteria for the geometrical search and the SCC iterations. The Monkhorst-Pack grid that converged the ground state energy was determined separately for each system; 96x96x1 for graphene, 16x16x1 for graphenylene, and 32x32x1 for each of the octafunctionalized biphenylenes (net-C, Type II, net-W).

Relative positions in a bilayer system are defined by the displacement vector,
\begin{equation}\label{eqn:D}
\mathbf{D}(a,b,h) = a \mathbf{A} + b \mathbf{B} + h \hat{z},
\end{equation}
where $\mathbf{A}$ and $\mathbf{B}$ are the in-plane lattice vectors, $a$ and $b$ form the ordered pair $[a,b] \in\mathbb{F}$, where $\mathbb{F}$ is given by the set theoretic equations
\begin{align}
\mathbb{A} &= \{a | a = 0.02n, n \in \mathbb{N}_{0} \wedge  n \le 50\}  \label{eqn:A}\\
\mathbb{B} &= \{b | b = 0.02n +0.01, n \in \mathbb{N}_{0} \wedge  n < 50\} \label{eqn:B} \\
\mathbb{F} &= \mathbb{A} \times \mathbb{A} \cup \mathbb{B} \times \mathbb{B} \label{eqn:F}
\end{align}
which parameterize the in-plane displacement, and
\begin{equation}\label{eqn:h}
h \in \{3.3, 3.325, 3.35, ..., 3.6\},
\end{equation}
is the out-of-plane displacement (interlayer separation) of one sheet with respect to the other.  In Equations \ref{eqn:A} and \ref{eqn:B} $\mathbb{N}_{0}$ are the natural numbers including zero, n is an element of the natural numbers less than (or equal to) 50, thus $a = 0.02n$ (or $b = 0.02n + 0.1$) when $n$ is defined as above.  At each displacement, $\mathbf{D}(a,b,h)$, the total energy was computed using a single point calculation. The single point calculations prevent unstable configurations from relaxing into more stable configurations.
The optimum interlayer spacing, $h_{opt}(a,b)$ was determined by fitting a Birch-Murnaghan equation of state (EOS) model to the interlayer spacing versus energy curve.  The curve fittings of the energy versus interlayer spacing had a maximum residual sum of squares value of 3.3e-06, demonstrating that the curves well describe the energies and interlayer spacings.  DFTB+ was then used to calculate band structures at each minimum energy displacement $\mathbf{D}(a,b,h_{opt}(a,b))$.  The out-of-plane bounding box was set to 30 \AA.

\section{Results and Discussion}

\begin{figure}
\centering
\includegraphics[clip,width=2.5 in, keepaspectratio]{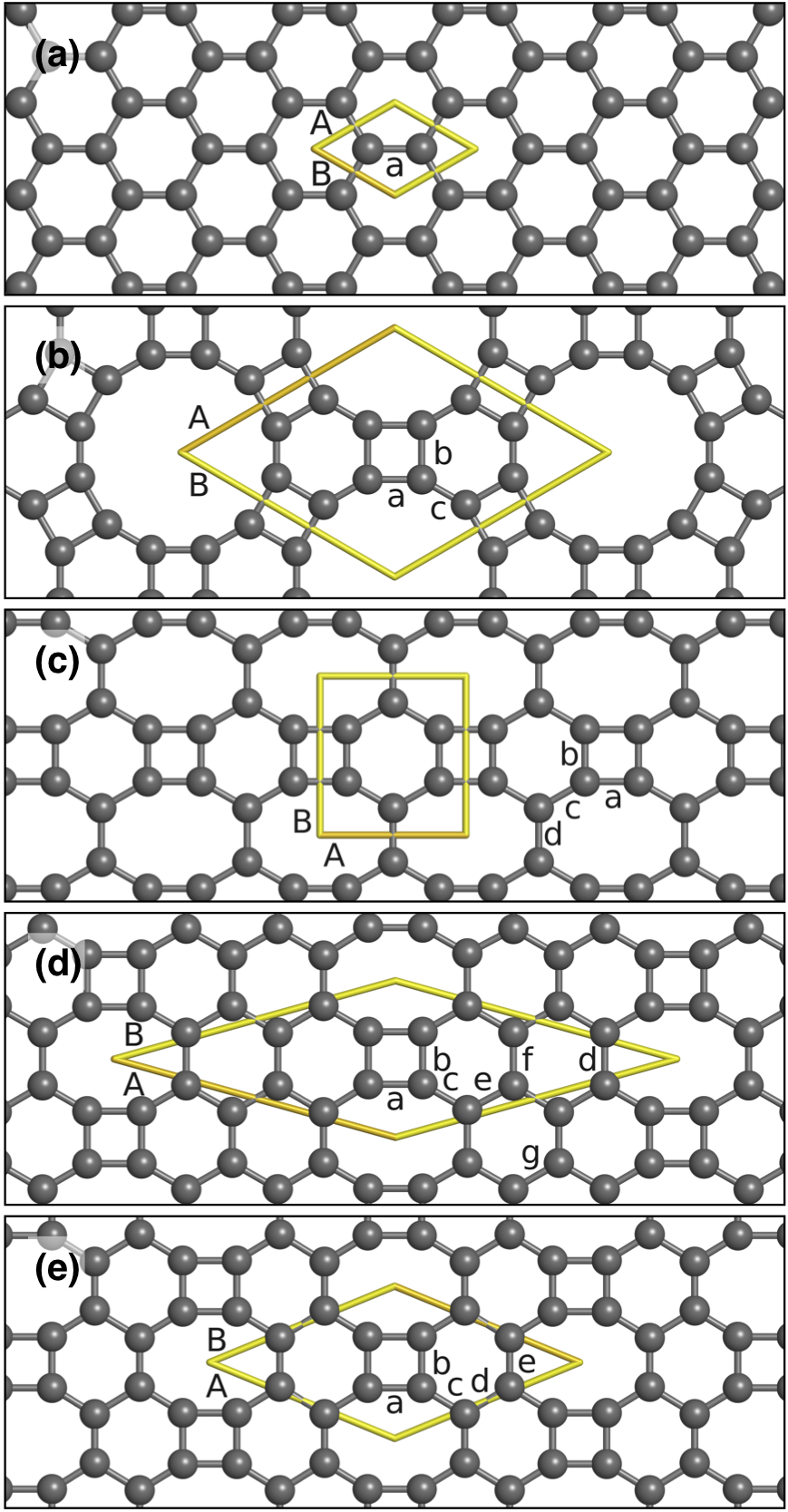}
\caption{Cell structures of (a) graphene, (b) graphenylene, (c) net-C, (d) Type II, (e) net-W. The gold colored parallelograms represent the edges of the 2-D unit cell, where A and B are the in-plane lattice vectors. Values for the bond lengths, a-g, are given in Table \ref{table:singlelayerbonds}. }
\label{fig:organiccells}
\end{figure}

\begin{figure}
\centering
\includegraphics[clip,width=3.4 in, keepaspectratio]{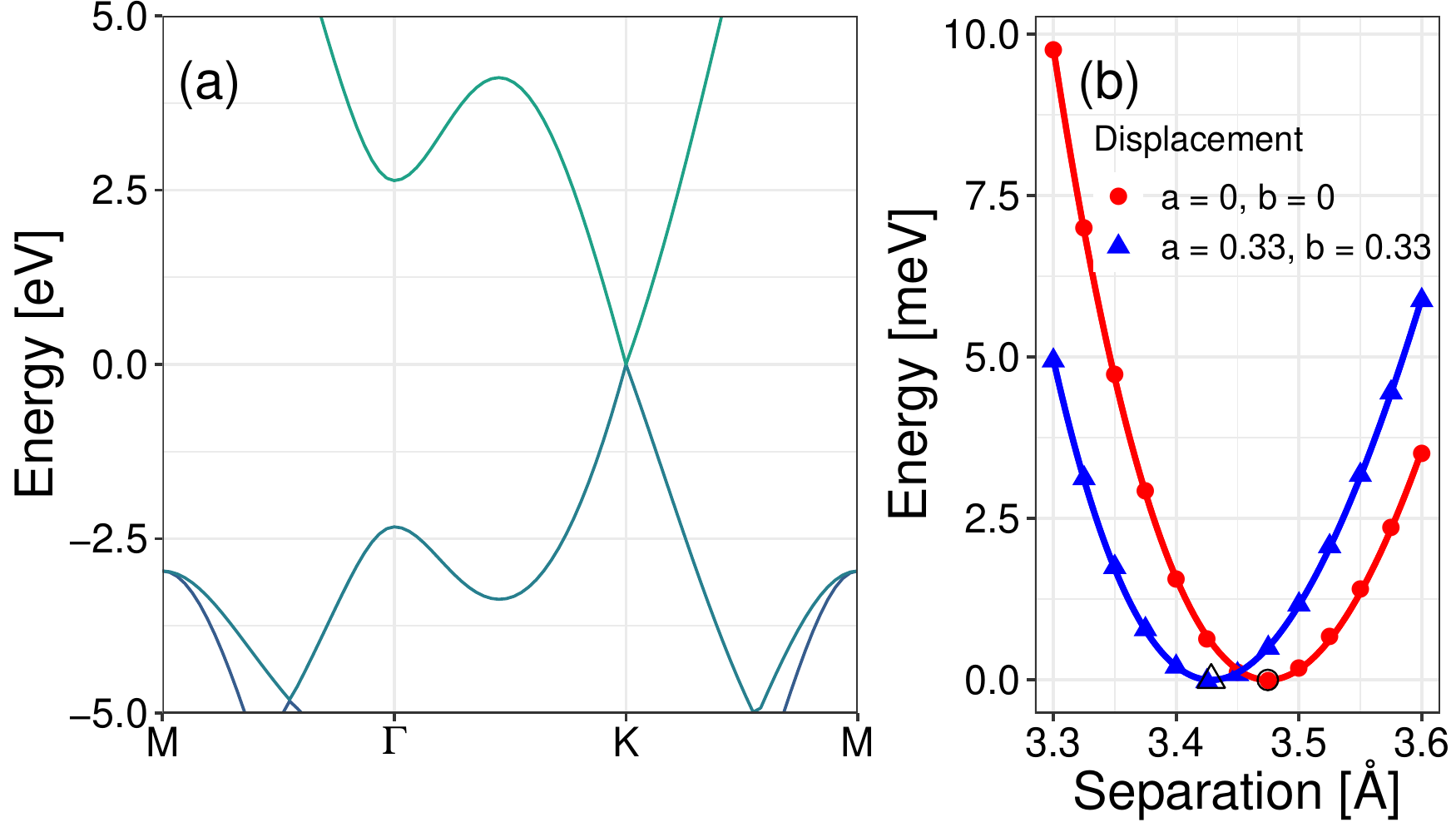}
\caption{(a) Band dispersion plot of graphene with energy zeroed at the Fermi level. (b) Energy versus interlayer separation for the in-plane displacements $[a,b]=[0,0]$ and $[0.33,0.33]$.  Red and blue points are the DFTB+ calculated results, lines are the respective fitted EOS, and black points are the optimal interlayer spacings given by the EOS.}
\label{fig:grbandplotEH}
\end{figure}

\begin{figure}
\centering
\includegraphics[clip,width=3.4 in, keepaspectratio]{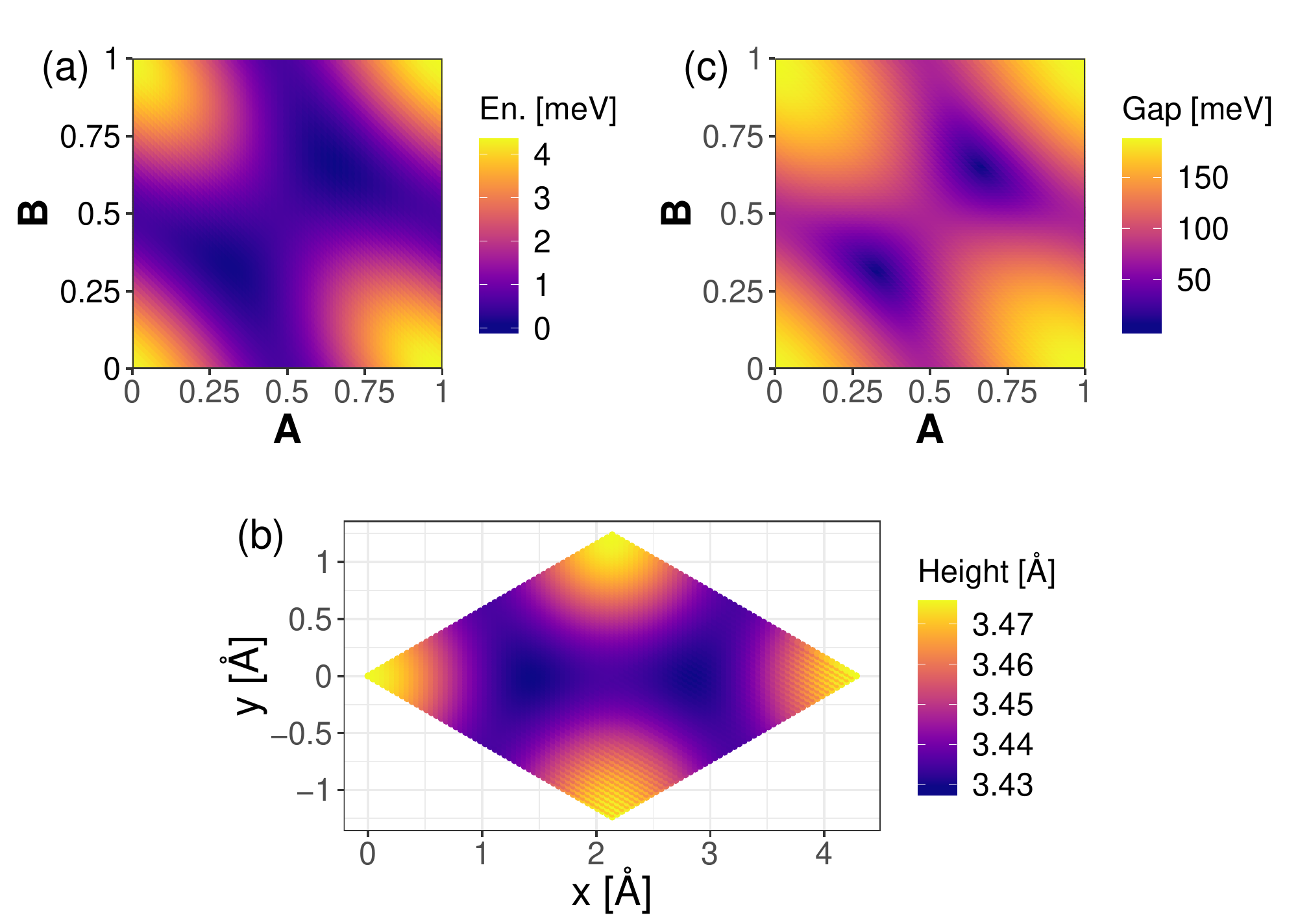}
\caption{(a) Relative total energy, (b) interlayer spacing, and (c) band gap of bilayer graphenes.}
\label{fig:graphene2D}
\end{figure}

As validation, the method was first used to study single layer and bilayer graphenes.  Figure \ref{fig:organiccells}(a) and Table \ref{table:singlelayerbonds} show that the C-C bond lengths and lattice vectors are in agreement with the commonly accepted values.   The present method gives a Dirac cone at the K high symmetry point in momentum space, see Figure \ref{fig:grbandplotEH} (a), consistent with previous studies \cite{CastroNeto2009}.  As further validation the method was used to scan the displacement space of bilayer graphene (BLG).    Figure \ref{fig:grbandplotEH} (b) presents energy versus interlayer spacing data for AA and AB stackings of graphene along with the fitted EOS.  Figures \ref{fig:graphene2D}(a) and (b) respectively show how the total energy and interlayer spacing change with displacement in the $\mathbf{A,B}$-plane as the top layer is displaced with respect to the bottom layer.  AA-stacking happens at $[a,b] = \{[0,0],\allowbreak [0,1],\allowbreak [1,0],\allowbreak [1,1]\}$ and AB-stacking $[a,b] = \{[1/3,1/3],\allowbreak [2/3,2/3]\}$.
The results show that AB-stacking has the lowest energy configuration and shortest interlayer spacing while AA-stacking presents the highest energy configuration and the largest interlayer spacing.
The ground state (AB-)stacking interlayer separation, 3.40 \AA, is consistent with other published theoretical values (3.38 \AA \cite{mostaani2015quantum}, 3.41 \AA \cite{zhang2013mechanical}, 3.4 \AA \cite{chakarova2006application}) and experiment (3.4 \AA \cite{ohta2006controlling}).  Figure \ref{fig:graphene2D}(c) shows that bilayer graphene is metallic at AB-stacking (consistent with Reference \cite{zhang2009direct}), and opens up to a gap of ~0.18 eV when at AA-stacking.

\begin{table*}[htb]
    \centering
\caption{Comparison of bond lengths, cell parameters, and total energies per carbon atom with published values.  The ground state of graphene is the zero of energies. Experimental values are in boldface text.}
\begin{scriptsize}
\begin{tabular}{cccccccccccccccccccc}
\hline
Value                    &    \multicolumn{2}{c}{graphene}    & &    \multicolumn{6}{c}{graphenylene}    &    &    \multicolumn{4}{c}{net-C}    & &    Type II    & &    \multicolumn{2}{c}{net-W}    \\
\cline{2-3}     \cline{5-10}     \cline{12-15}     \cline{17-17}     \cline{19-20}\\
    &    Current    &    \citen{bosak2007elasticity}    &    &    Current    &    \citen{du1740796}    &    \citen{Song1338}     &    \citen{Brunetto1212810}     &    \citen{C6TA04456E}     &    \citen{de2013ab}     &    &    Current    &    \citen{Rahaman201765}     &    \citen{karaush2016computational}     &    \citen{Ferguson201720577}     &    &    Current    &    &    Current    &    \citen{wang2013prediction}     \\
a [\AA]                  &    1.43    &        &    &    1.51    &    \textbf{1.50$\thicksim$1.52}    &    1.48    &    1.49    &    1.48    &    1.48    &    &    1.54    &    1.45    &    1.45    &        &    &    1.54    &    &    1.52    &        \\
b [\AA]                  &        &        &    &    1.48    &    \textbf{1.42$\thicksim$1.46}    &    1.47    &    1.48    &    1.47    &    1.47    &    &    1.43    &    1.46    &    1.46    &        &    &    1.42    &    &    1.44    &    1.48    \\
c [\AA]                  &        &        &    &    1.37    &    \textbf{1.42$\thicksim$1.46}    &    1.37    &    1.37    &    1.37    &    1.36    &    &    1.41    &    1.41    &    1.41    &        &    &    1.39    &    &    1.4    &    1.44    \\
d [\AA]                  &        &        &    &                      &        &               &        &        &        &    &    1.47    &    1.45    &    1.44    &        &    &    1.45    &    &    1.46    &    1.39    \\
e [\AA]                  &        &        &    &                      &        &               &        &        &        &    &                  &        &        &        &    &    1.44    &    &    1.46    &    1.46    \\
f [\AA]                  &        &        &    &                      &        &               &        &        &        &    &                  &        &        &        &    &    1.44    &    &        &    1.45    \\
g [\AA]                  &        &        &    &                      &        &        &        &        &        &    &                  &        &        &        &    &    1.44    &    &        &        \\
$|\mathbf{A}|$ [\AA]     &    2.47    &    \textbf{2.46}    &    &    6.83    &        &    6.76    &    6.74    &    6.72    &    6.75    &    &    7.96    &                &    3.75    &    3.78    &    &    15.38    &    &    5.62    &    5.49    \\
$|\mathbf{B}|$ [\AA]     &    2.47    &    \textbf{2.46}    &    &    6.83    &        &    6.76    &    6.74    &    6.72    &               &    &    4.31    &                &    4.52    &    4.53    &    &    4.26    &    &    5.62    &    5.49    \\
$\angle$ [deg]             &    30    &    \textbf{30}    &    &    30    &        &        &    30    &    30    &    30    &    &    90    &                &               &    90    &    &    30.96    &    &    44.88    &    47.7    \\
En [eV/C]    &    0    &        &    &    0.49    &        &    0.66    &    0.63    &        &    0.66    &    &    0.36    &    0.46    &        &        &    &    0.05    &    &    0.22895    &    0.37    \\
\hline
    \end{tabular}
\end{scriptsize}
    \label{table:singlelayerbonds}
\end{table*}

\begin{figure}
\centering
\includegraphics[clip,width=3.4 in, keepaspectratio]{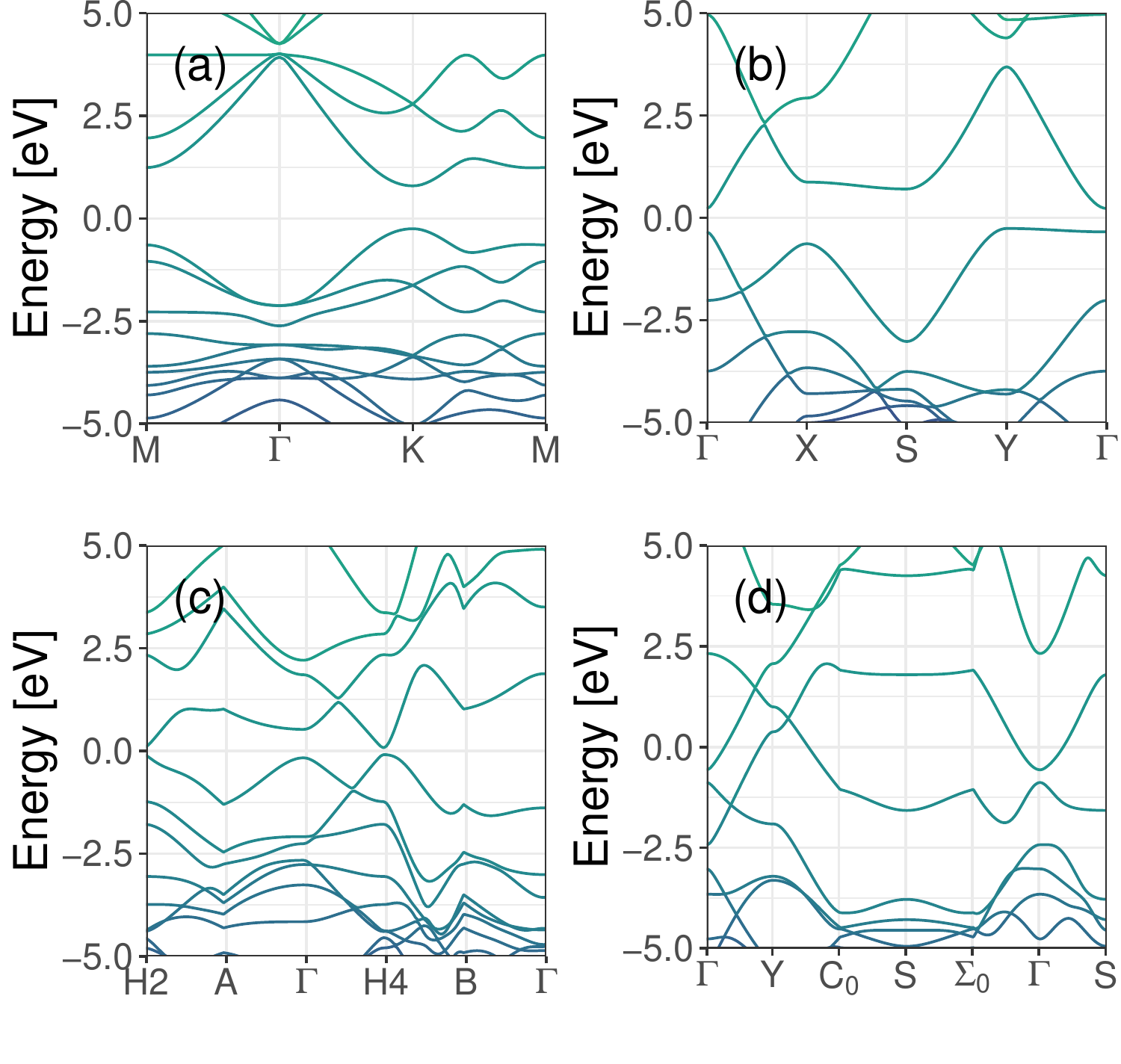}
\caption{Band dispersion plots of (a) graphenylene, (b) net-C, (c) Type II, and (d) net-W.}
\label{fig:SLband}
\end{figure}

Figure \ref{fig:organiccells} and Table \ref{table:singlelayerbonds} contain the presently computed bond lengths and lattice constants of single layers of graphenylene, net-C, Type II, and net-W which compare well with published values.  Band dispersion plots of these structures are presented in Figure \ref{fig:SLband}.  For single layer graphenylene this method results in a band gap of 1.04 eV, which is significantly larger than values given by DFT with GGA functionals (i.e. 0.034 eV \cite{Lu133677}, 0.025 eV \cite{Song1338}, 0.033 eV \cite{koch2015graphenylene}) but only three percent smaller than the value given by DFT with the hybrid functional B3LYP, 1.08 eV \cite{de2013ab}.  The band gap values reported in the literature are expected because DFT is known to underestimate band gap values when using either LDA or GGA functionals and hybrid functionals generally find band gap values close to experimental values.  DFTB+ often gives results that are near experimental values, in part, because the Fermi compression in solids is compensated by the excited nature of compressed spherical orbitals \cite{haycock2011calculation}.  With the present method Net-C has a band gap of 0.488 eV, much lower than the 2 eV band gap found by Tyutyulkov \textit{et al.} and contrary the metallic band structure found by Hudspeth \textit{et al.} \cite{tyutyulkov1997structure, hudspeth2010electronic}. Type II has a band gap of 0.17 eV. While net-W is conducting in nature, as was found by Wang \textit{et al.} \cite{wang2013prediction} but contrary to the zero gap semiconductor found by Xu \textit{et al.} \cite{Xu141113}.

\begin{figure}
\centering
\includegraphics[clip,width=3.4 in, keepaspectratio]{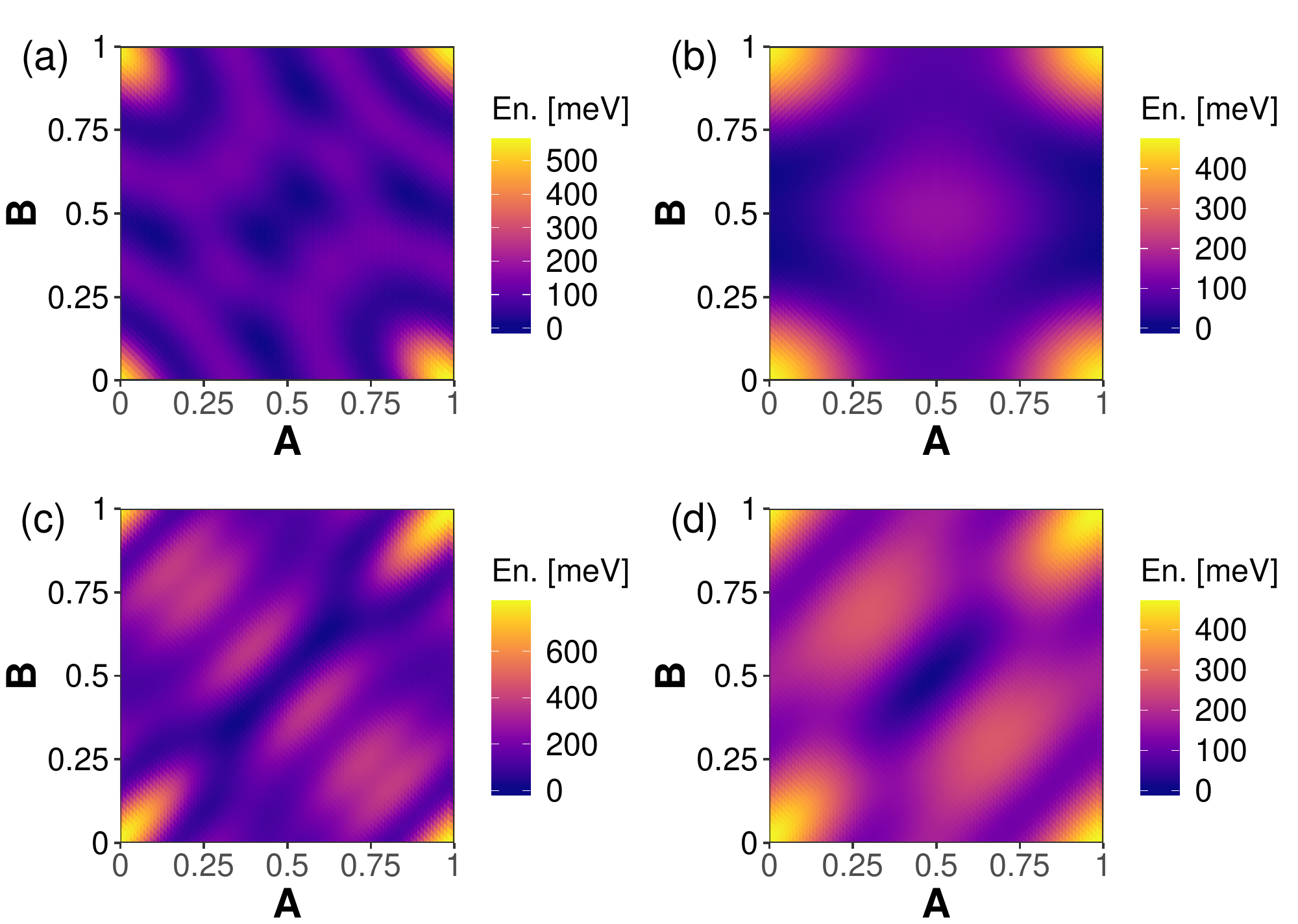}
\caption{Relative total energies of in-plane displacements of bilayer (a) graphenylene, (b) net-C, (c) Type II, and (d) net-W.}
\label{fig:2DrelativeEnergies}
\end{figure}

The graphs in Figure \ref{fig:2DrelativeEnergies} show how the relative total energies change with with $[A,B,h_{opt}(a,b)]$-displacement for the biphenylene bilayers.
The optimum interlayer spacing $h_{opt}(a,b)$ at each point $[a,b]$ was found according to the method described in the discussion of Equation \ref{eqn:D}, and illustrated for two graphene displacements in Figure \ref{fig:grbandplotEH} (b).
For each system AA stacking is the overall maximum energy position-an unstable equilibrium position. For graphenylene minimum energy positions are near in-plane displacements $[a,b] = \{[0.13, 0.43],\allowbreak [0.43, 0.13],\allowbreak  [0.57, 0.87],\allowbreak [0.87, 0.57]\}$.  For net-C the range in relative total energies is 17 meV, with the minimum energy positions at the in-plane displacements $[a,b] = \{[0.0, 0.4],\allowbreak [0.0, 0.6],\allowbreak [1.0, 0.4],\allowbreak [1.0, 0.6]\}$. Note that points $[0.0, 0.4]$ and $[1.0, 0.4]$, as well as $[0.0, 0.6]$ and $[1.0, 0.6]$, are invariant under translational symmetry.  According to Ferguson \textit{et al.} the AA stacking should be the most stable form of bilayer net-C \cite{Ferguson201720577}.
The range in relative total energies, 29 meV, is largest for Type II bilayers.  For these systems the minimum energy in-plane displacements are at $[a,b] = \{[0.36, 0.36],\allowbreak [0.64, 0.64]\}$.
The minimum energy in-plane displacement is at $[a,b] = \{[0.50, 0.50]\}$ for bilayer net-W, coming from a range in total energies of 17 meV. Computing the relative energy per unit area for each bilayer structure discussed in this work shows that the greatest change in energy per unit area happens with BLG, likely because BLG has more atoms per unit area, thus larger steric effects per unit area.

\begin{figure}
\centering
\includegraphics[clip,width=3.4 in, keepaspectratio]{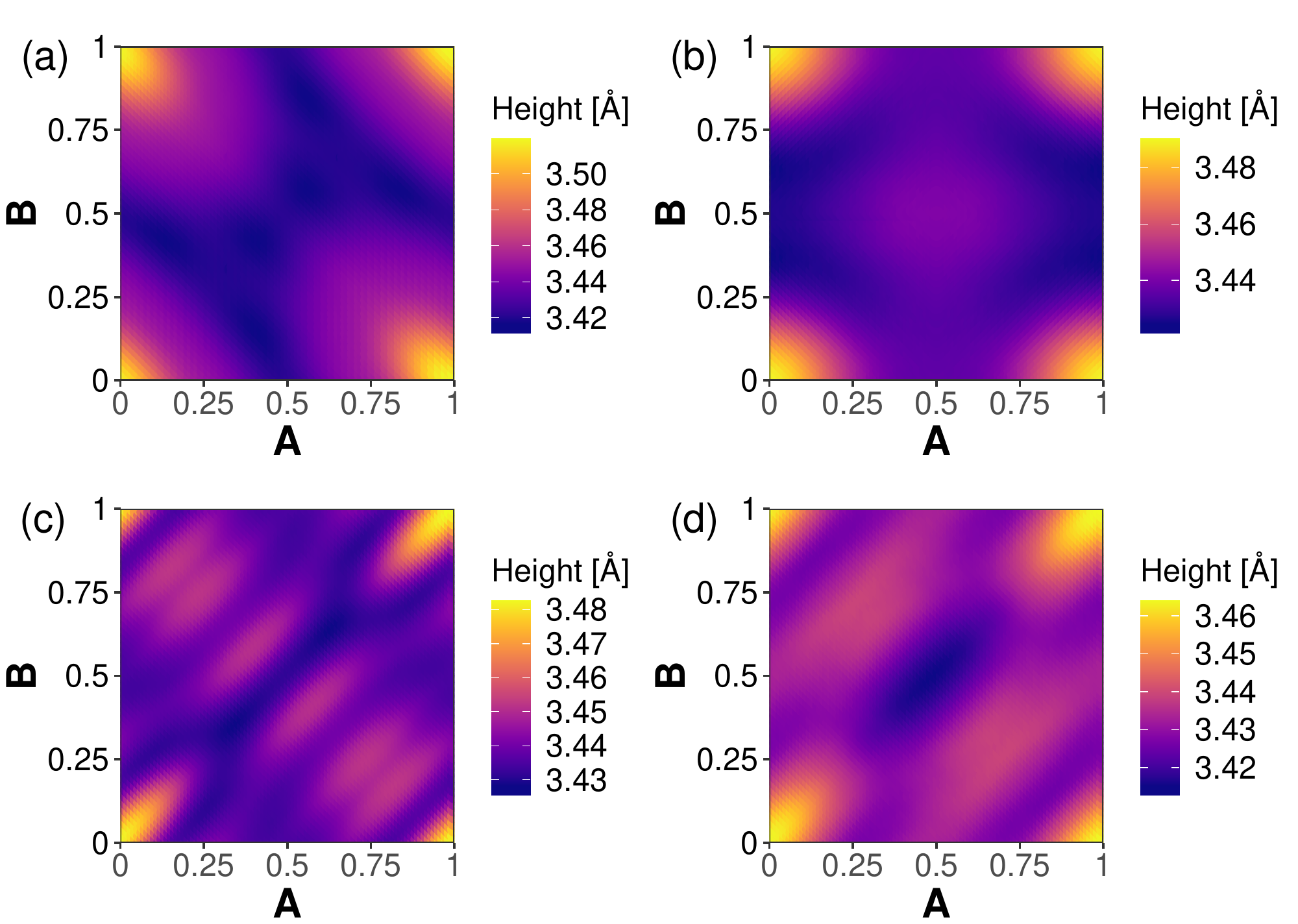}
\caption{Interlayer spacing of in-plane displacements of bilayer (a) graphenylene, (b) net-C, (c) Type II, and (d) net-W.}
\label{fig:XYheights}
\end{figure}

Figure \ref{fig:XYheights} presents the optimum interlayer spacing, $h_{opt}$, of each bilayer given with respect to $[a,b]$-displacement.
For graphenylene, the interlayer spacing at all of the minimum energy displacements is 3.41 \AA.  These points also correspond to the minimum interlayer spacing for graphenylene.  The maximum interlayer spacing for graphenylene happens at AA stacking with $h_{opt} = 3.48$ \AA.  The minimum energy displacements for net-C have interlayer spacings of $h_{opt} = 3.42$ \AA.
Contrary to what we see in BLG, the minimum energy position does not always correspond to the minimum interlayer spacing.  Bilayers of graphenylene, net-C, and net-W all have a predicted minimum interlayer spacing at a point other than the minimum energy position; though they are within 0.03 lattice constant of the energy minimum.  This difference between the minimum interlayer spacing and minimum energy positions might be due to small errors associated with the fitting process or may be due to the coarse nature of the AB plane parameterization not finding the actual minimum positions.

\begin{figure}
\centering
\includegraphics[clip,width=3.4 in, keepaspectratio]{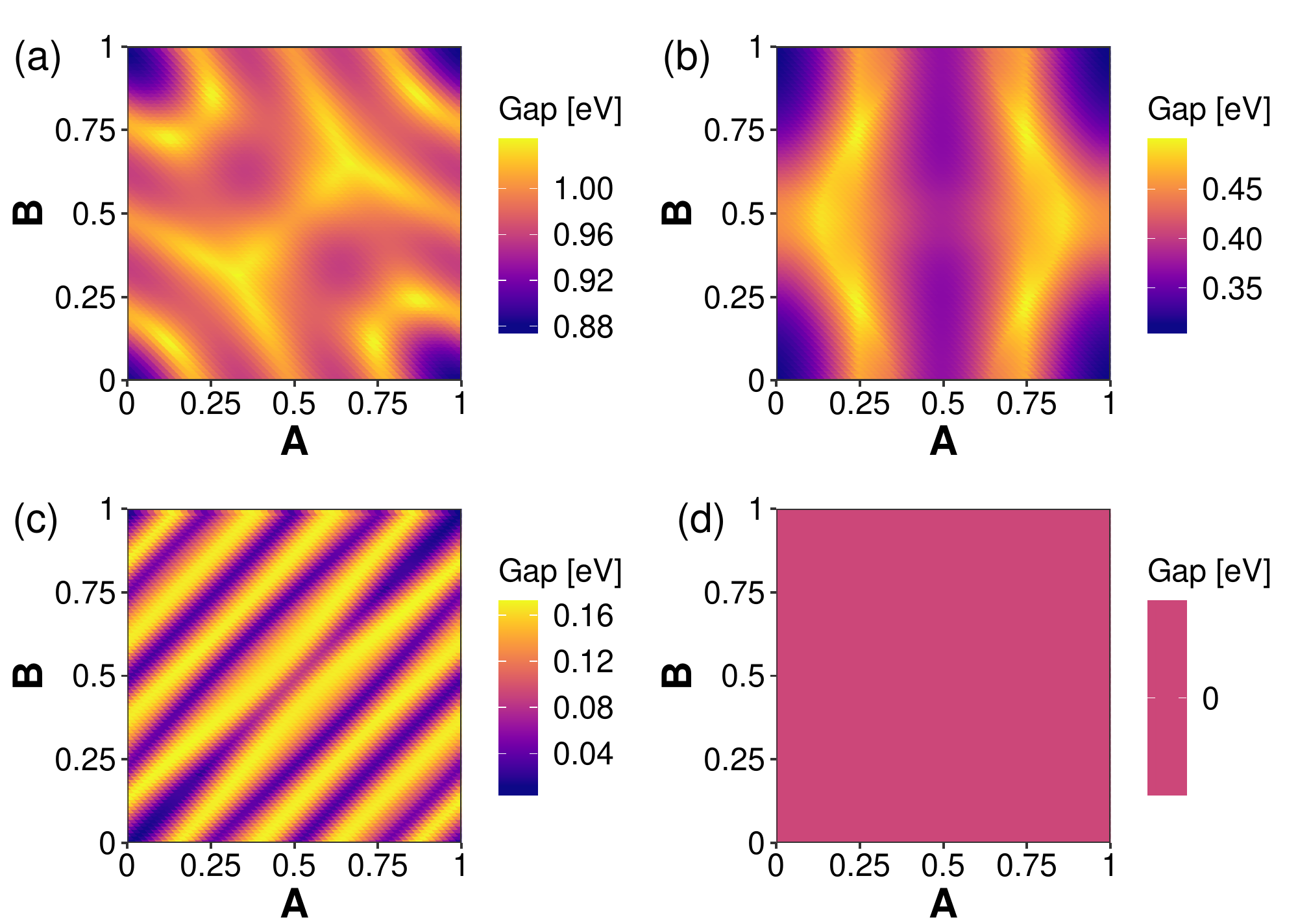}
\includegraphics[clip,width=3.4 in, keepaspectratio]{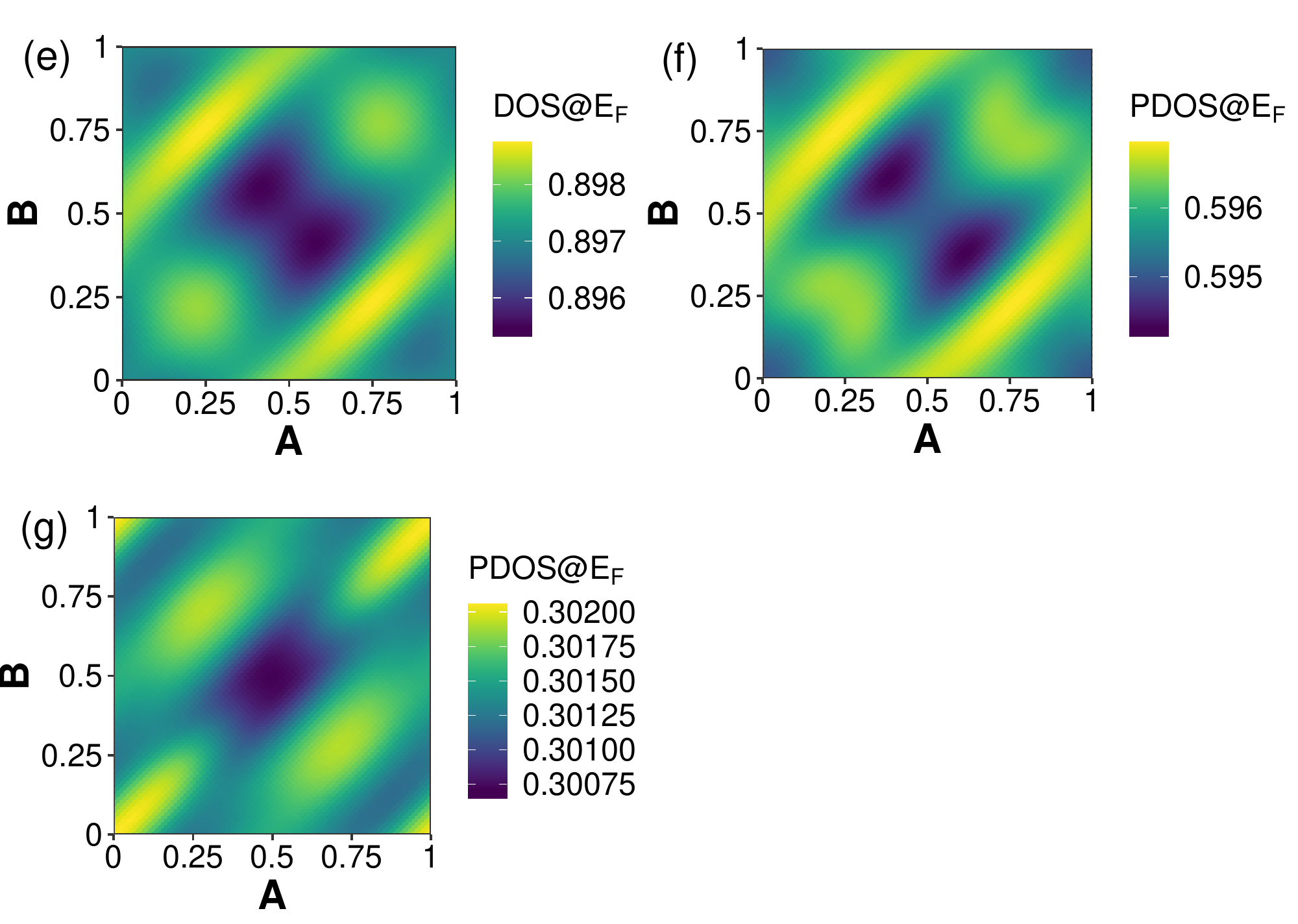}
\caption{Band gaps of in-plane displacements of bilayer (a) graphenylene, (b) net-C, (c) Type II, and (d) net-W.  (e) DOS, (f) PDOS 4-member ring p-orbitals, (g) PDOS other atoms p-orbitals at the Fermi level for in-plane displacements of bilayer net-W.}
\label{fig:2Dbandgaps}
\end{figure}

Figure \ref{fig:2Dbandgaps} details how electronic structure around the Fermi energy changes for each bilayer as one layer is displace from the other by $[a,b, h_{opt} (a,b)$.
Bilayer graphenylene band gaps, presented in Figure \ref{fig:2Dbandgaps}(a), range from 0.878 eV to 1.04 eV with the minimum energy positions having a band gap of 1.02 eV.  The band gaps of bilayer net-C, Figure \ref{fig:2Dbandgaps}(b), range from 309 meV to 496 meV, with a band gap of 424 meV at the minimum energy positions.  These results demonstrate that it is possible to use these materials to create a device that has a reliable band gap without BLG's need for ensuring a proper stacking configuration. The band gaps of bilayer Type II, Figure \ref{fig:2Dbandgaps}(c), range from 7.94 meV to 168 meV, with a band gap of 60.2 meV at the minimum energy positions. Bilayer net-W, Figure \ref{fig:2Dbandgaps}(d), is conducting for in-plane displacements.  Again, this demonstrates that we could produce a bilayer that is conducting without needing to ensure a particular stacking configuration.

For conductors, such as net-W, the density of states (DOS) at the Fermi energy is a significant quantity used in computing many materials properties, such as heat capacity and spin susceptibility \cite{kenner1975surface, batabyal2014evolution}.
The DOS and the atomic orbital resolved partial DOS (PDOS) of the in-plane displacements of net-W were computed.  The PDOS was computed for the s- and p-orbitals of the 4-member ring carbons, and then the rest of the carbon atoms. The estimated DOS (and PDOS) at the Fermi level was computed by linear interpolation.  Figure \ref{fig:2Dbandgaps}(e-g)  shows the results of the linear interpolation at each in-plane displacement.  At the Fermi level, the values of the s-orbital PDOS are five orders of magnitude smaller than the p-orbital PDOS.  The DOS of states ranges from 0.895 to 0.899 states/eV, with the maximum (minimum) DOS at $[a,b] = \{[0.24, 0.76],\allowbreak [0.76, 0.24]\}$ ($[a,b] = \{[0.42, 0.58],\allowbreak [0.58, 0.42]\}$). At the minimum energy position, the DOS has a value of 0.896 states/eV.
Nearly two-thirds of the DOS at the Fermi level is made up of 4-member ring p-orbitals while the rest is made up almost entirely of the other carbon atoms in the unit cell (which are part of 6-member rings) p-orbitals.  The PDOS give maximum and minimum values at different in-plane displacements for the 4- and 6-member rings, with the maximum (minimum) of the DOS being in nearly the same position as the 6-member ring PDOS maximum (minimum) position.

\begin{figure}
\centering
\includegraphics[clip,width=3.4 in, keepaspectratio]{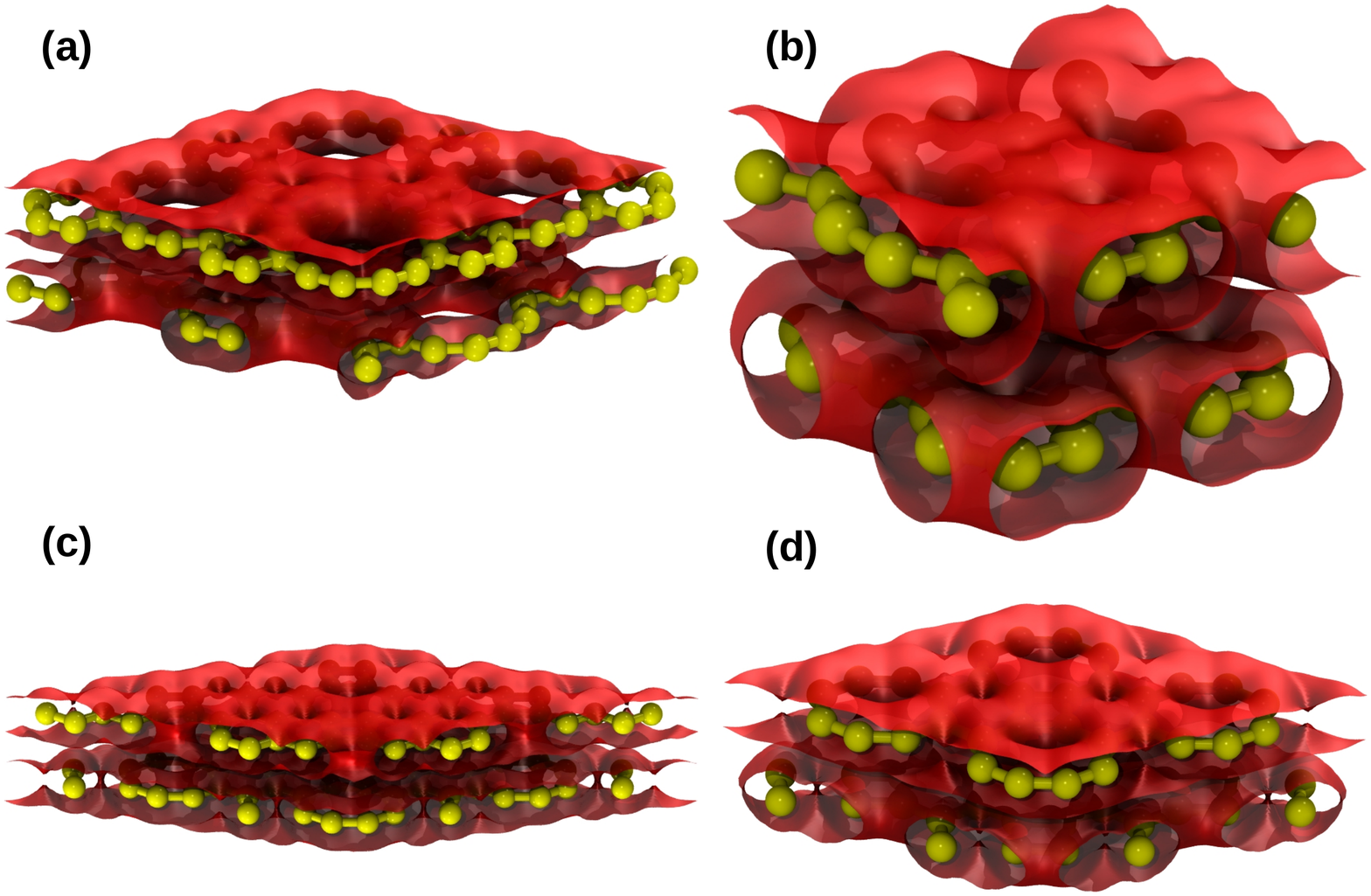}
\caption{3D view of total charge density for (a) Graphenylene, (b) net-C, (c) Type II and (d) net-W. The plotted isosurfaces refer to a constant electron density of 0.01. }
\label{fig:charge_iso}
\end{figure}

\begin{figure*}
\centering
\includegraphics[clip,width=5 in, keepaspectratio]{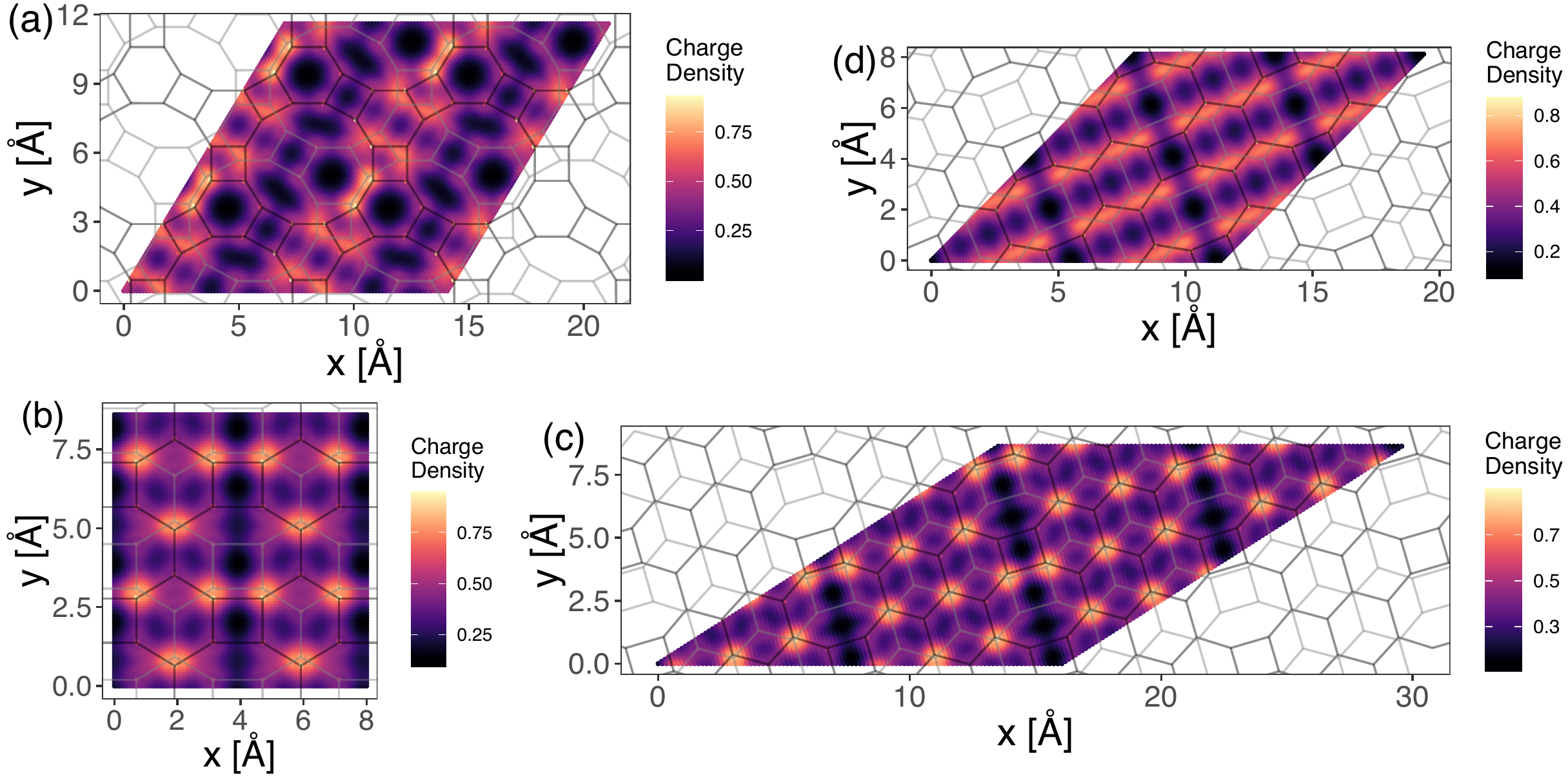}
\caption{2D plots of the total charge density for (a) Graphenylene, (b) net-C, (c) Type II and (d) net-W. The plotted map refers to the integration of the charge density along the z direction in the region of the two layers.}
\label{fig:totalcharge}
\end{figure*}

\begin{figure*}
\centering
\includegraphics[clip,width=5 in, keepaspectratio]{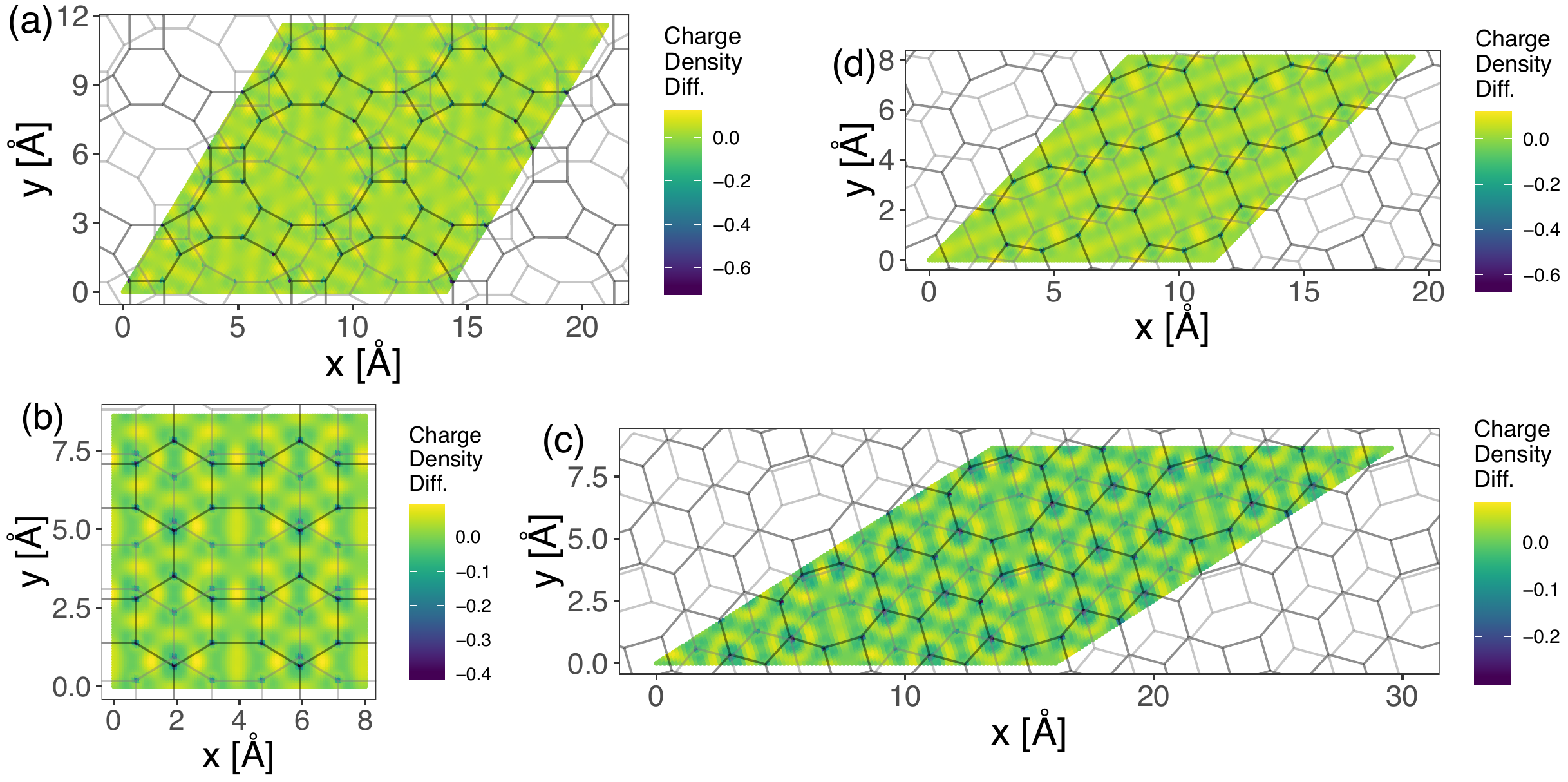}
\caption{2D plots of total charge density difference for (a) graphenylene, (b) net-C, (c) Type II and (d) net-W. Plotted maps refer to the difference between the charge value that would be found considering a superposition of neutral atoms and the actual charges found after the SCC calculation.}
\label{fig:chargediff}
\end{figure*}

Figure \ref{fig:charge_iso} displays the electron charge density isosurfaces at the constant value $\rho_c=0.01$ ($\rho_c=1.0$ would indicate the charge of one electron) for (a) graphenylene, (b) net-C, (c) Type II and (d) net-W supercells. In the before-mentioned figure, one can see that, in all cases, the presence of regions with no charge, forming pores indicating the electrostatic effect imposed by the underlying porous atomic lattices. Particularly, it is interesting to note that graphenylene and net-W lattices present large pores quite close to each other while net-C and Type II bilayers have smaller pores and bigger density of small holes on the surface.
The geometrical configuration of the {\sl electrostatic} pores in these materials will be relevant in their use in molecular docking and molecular separation.
Figure \ref{fig:totalcharge} shows 2-dimensional maps for the four structures cited above. In these maps, for each position in a plane parallel to the lattice, the density was integrated in the z-(i.e., perpendicular) direction in the region of the bilayer. These maps show a higher electron density in a few regions, especially when there are bonds crossing. Dark regions indicate low or null densities and are noted mostly in the presence of superposed pores of both layers.
On the other hand, Figure \ref{fig:chargediff} shows 2-dimensional maps of the calculated charge density difference between the value that would be found considering a superposition of neutral atoms and the actual charges found after the SCC calculation. The presented results were also integrated along z-direction for each point. This calculation indicates approximately the charge exchanges between different parts of the atomic structure. Our results suggest a tendency of the electrons of the system to transfer from carbon atoms to regions between adjacent atoms, where the shared pairs of electrons should be most of the time in covalent bonds.

\begin{figure}
\centering
\includegraphics[clip,width=3.4 in, keepaspectratio]{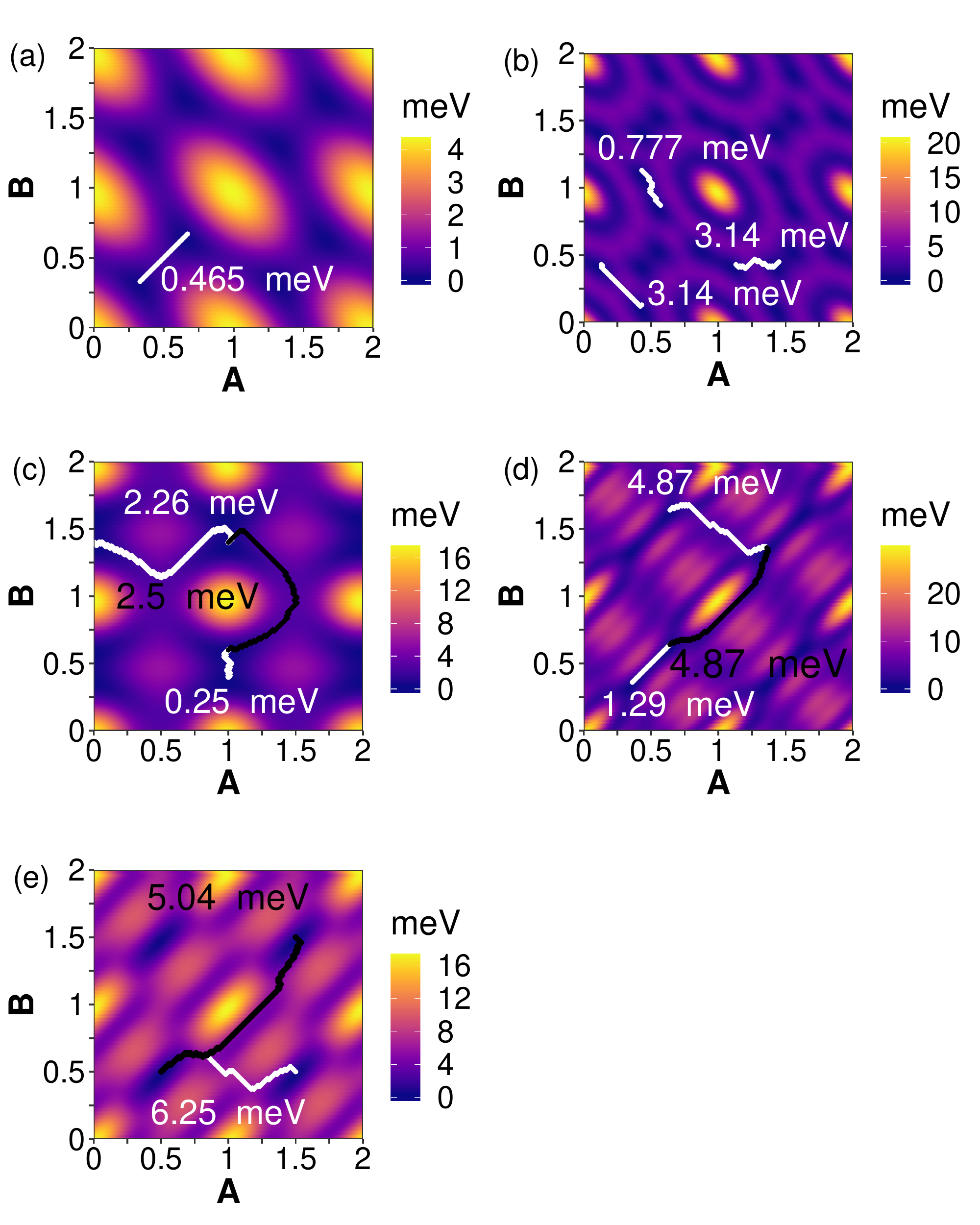}
\caption{Minimum energy paths between ground state energy points in two-by-two displacement-space supercells. Paths are designated with white (or black) lines and are labeled with the maximum relative energy of the path.}
\label{fig:minimumenergypaths}
\end{figure}

Figure \ref{fig:minimumenergypaths} gives the minimum energy stacking fault paths between pairs of ground state displacements.  The paths were determined using network analysis where the nodes of the graph are the displacements described in Equation \ref{eqn:D}. The cost along each edge is the absolute value of the change in energy between connected nodes.  A breadth-first search function was used to find the path with the lowest total cost \cite{ubergraph2018}.  In bilayer graphene, there are two minima energy positions in a unit cell, at AB and AB' stacking. Figure \ref{fig:minimumenergypaths} (a) shows four unit cells of bilayer graphene in displacement space. In the graphitic systems the paths between all ground state positions are invariant under translational and rotational symmetry; thus only one path is shown.  In this case, an energy barrier between neighboring AB-stacking positions has a height of 0.48 meV.
Bilayer graphenylene paths were computed for three pairs of energy minima. All three bilayer paths have energy barriers that are larger than the BLG energy barrier.  Except for one very short path in net-C systems, the energy barriers of all the biphenylene systems are significantly larger than in BLG.  The greater barriers indicate that these systems undergo sheering motion less readily than graphitic systems.

\section{Summary}

The possible device applications of these bilayers are manifold.
For example, replacing multilayer graphene oxide, which is known to degrade with time \cite{kim2012room}, with layers of the more inert graphenylene (or net-C) in flexible thin film transistor \cite{SeoungKiLee20123472}.
The conducting nature of bilayer (or few-layer) net-W  may find use in flexible transparent conductors \cite{C1NR11574J,Parveznn400576v}, anodes for novel alkali (and alkaline earth) metal-based fuel cells\cite{Sun2019acsaem}, and in emerging battery technologies \cite{Mukherjee8b00843,Song2018NetWLiion}.  In many cases, the porous nature of the carbon allotropes discussed in this work is a desirable feature in either the device processing or in the device itself.  For example, depending on the degree of strain, lithium and some small molecules have been shown to propagate through a graphenylene surface \cite{C6TA04456E,koch2015graphenylene}.  While this porous nature has been studied in terms of gas separation and lithium adsorption, it also means that layers of a material could be easily doped.

Industrial scale device fabrication requires materials with reliable electronic properties. This work found three bilayer systems with stable electronic properties under in-plane displacement.  Bilayers of graphenylene and net-C are semiconducting for all sheer displacements, and net-W is conducting for all sheer displacements.  Whereas, bilayer Type II, like BLG, is conducting or semiconducting depending on the sheer displacement.  The method used for our calculations correctly predicted bond lengths, unit cell dimensions, and band dispersion of both SLG and of the biphenylenes that have been studied previously. It further demonstrates that the method correctly predicts that AB stacking is the ground state of BLG and gives an interlayer separation that is consistent with previous studies.  Thus, this work demonstrated that three of the materials studied here are reliably either conducting or semiconducting as bilayers making them good candidates for device fabrication.

Further work could include the effects of rotation on band structures \cite{Kolmogorov05235415,wang2012interfacial,zhang2013mechanical}, computing elastic constants \cite{savini2011bending}, the modeling of the effect of electrical bias on the bilayers \cite{oostinga2008gate}, or computing thermal conductivities \cite{Lindsay205441}.

\section*{Acknowledgement}

The production of atomic structure models and data analysis was performed using Greenwood: A library for creating molecular models and processing molecular dynamics simulations \cite{greenwood2018}. Further data analysis was performed using R: A language and environment for statistical computing \cite{RcoreTeam2014, Wickham2009, Wickhamdplur2016, Dowledatatable, Garnier2017viridis} and Ubergraph \cite{ubergraph2018}. Seek-K-path was used to determine the high symmetry kpoints of systems \cite{HINUMA2017140} and to verify crystal space groups.

\section*{Author Contributions}
C.E.J. and R.P. conceived the project, discussed the results, and wrote the manuscript.  C.E.J. carried out the experiments, performed needed programming, analyzed the results.  R.P. provided computing resources.

\section*{Data availability}

The raw data required to reproduce these findings are available to download from [http://dx.doi.org/10.17632/ytpycsd6ww.1]. The processed data required to reproduce these findings are available to download from [http://dx.doi.org/10.17632/ytpycsd6ww.1].



\bibliographystyle{elsarticle-num.bst}
\bibliography{bilayer}





\end{document}